\documentclass[12pt,a4paper]{article}
\usepackage[english,french]{babel}
\usepackage[utf8]{inputenc}
\usepackage{fontawesome}
\usepackage{a4wide}
\usepackage{amsmath}
\usepackage{amsthm}
\usepackage{amssymb}
\usepackage{cleveref}
\usepackage{stmaryrd}
\usepackage{dsfont}
\usepackage{tikz}
\usepackage{xcolor}
\usepackage{svg}
\usepackage{soulutf8}
\usepackage{rotating}
\usepackage[affil-it]{authblk}
\definecolor{lviolet}{rgb}{.95,.85,1}
\definecolor{lblue}{rgb}{.9,.9,1}
\definecolor{lred}{rgb}{1,.85,.85}
\definecolor{lgreen}{rgb}{.85, 1,.85}
\definecolor{lgray}{rgb}{.92,.92,.92}
\definecolor{lorange}{rgb}{1,.92,.8}

\theoremstyle{definition}
\newtheoremstyle{break}
{\topsep}{\topsep}%
{\itshape}{}%
{\bfseries}{}%
{\newline}{}%
\theoremstyle{break}
\newtheorem{definition}{Définition}
\newcommand\blfootnote[1]{%
	\begingroup
	\renewcommand\thefootnote{}\footnote{#1}%
	\addtocounter{footnote}{-1}%
	\endgroup
}

\makeatletter
\date{11 octobre 2018}
\DeclareMathOperator{\argmax}{argmax}

\title{MoCaNA, un agent de négociation automatique utilisant la recherche arborescente de Monte-Carlo}
\author[1,2]{Cédric L.R. Buron\thanks{cedric.buron@thalesgroup.com}} %
\author[1,3]{Zahia Guessoum\thanks{zahia.guessoum@lip6.fr}}
\author[4]{Sylvain Ductor\thanks{sylvain.ductor@uece.br}}
\author[5]{Olivier Roussel\thanks{olivier.roussel@kyriba.com}}
\affil[1]{LIP6, Sorbonne Université, Paris, France}
\affil[2]{Thales Research and Technology, Palaiseau, France}
\affil[3]{CReSTIC, Université de Reims, Reims, France}
\affil[4]{Universidade Estadual do Cear\'a, Fortaleza, Br\'esil}
\affil[5]{Kyriba Corp, San Diego, USA}

\begin{document}
	\selectlanguage{french} 
	\maketitle
	\blfootnote{Vingt-sixièmes Journées Francophones sur les Systèmes Multi-Agents (JFSMA'18), Métabief, France, p. 85--94, oct. 2018}
		
	\begin{abstract}
		La négociation automatique est un sujet qui suscite un intérêt croissant dans la recherche en IA. Les méthodes de Monte-Carlo ont quant à elles vécu un grand essor, notamment suite à leur utilisation sur les jeux à haut facteur de branchement tel que le go. 
		
		Dans cet article, nous décrivons un agent de négociation automatique, Monte-Carlo Negotiating Agent (MoCaNA) dont la stratégie d'offre s'appuie sur la recherche arborescente de Monte Carlo. Nous munissons cet agent de méthodes de modélisation de la stratégie et de l'utilité adverse. MoCaNA est capable de négocier sur des domaines de négociation continus et dans un contexte où aucune borne n'est spécifiée. Nous confrontons MoCaNA aux agents de l'ANAC 2014 et à un \textsl{RandomWalker} sur des domaines de négociation différents. Il se montre capable de surpasser le \textsl{RandomWalker} dans un domaine sans borne et la majorité des finalistes de l'ANAC dans un domaine avec borne.
		
		\noindent\textbf{Mots-clés:} Monte-Carlo, Négociation automatique, Agent
	\end{abstract}
	\selectlanguage{english} 
	\begin{abstract}
		\vspace{.5cm}	
		Automated negotiation is a rising topic in Artificial Intelligence research. Monte Carlo methods have got increasing interest, in particular since they have been used with success on games with high branching factor such as go.
		
		In this paper, we describe an Monte Carlo Negotiating Agent (MoCaNA) whose bidding strategy relies on Monte Carlo Tree Search. We provide our agent with opponent modeling tehcniques for bidding strtaegy and utility. MoCaNA can negotiate on continuous negotiating  domains and in a context where no bound has been specified. We confront MoCaNA and the finalists of ANAC 2014 and a RandomWalker on different negotiation domains. Our agent ouperforms the RandomWalker in a domain without bound and the majority of the ANAC finalists in a domain with a bound.
		
		\noindent\textbf{Keywords:} Monte Carlo, Automated Negotiation, Agent
	\end{abstract}
	\selectlanguage{french} 
%

	\section{Introduction}
	
	La négociation est une forme d'interaction dans laquelle un groupe d'agents ayant des conflits d'intérêt et un désir de coopérer essaient de trouver un accord mutuellement acceptable sur un objet de négociation. Ils explorent à cette fin les solutions selon un protocole prédéfini.
	
	La question de l'automatisation de la négociation, bien connue dans les domaines économiques depuis l'avènement des applications de e-commerce, a reçu une attention toute particulière dans le champ de l'intelligence artificielle et des systèmes multi-agents.
	
	De nombreux \textsl{frameworks} ont été proposés \cite{Sandholm1999Distributed}: enchères, réseaux contractuels, équilibres généraux; ils permettent de négocier selon plusieurs modalités. Les négociations peuvent être caractérisées selon divers aspects portant notamment sur l'ensemble des participants (bilatéral ou multilatéral), les préférences des agents (linéaires ou non), les attributs sur lesquels la négociation porte (discrets ou continus), ou encore les caractéristiques du protocole (borné ou non par le temps ou le nombre de tours). Chaque agent utilise une stratégie pour évaluer l'information reçue et faire des offres. Plusieurs stratégies ont été proposées comme \cite{Faratin1998Negotiation,Jonge2016,Szoellosi2016}. Elles peuvent être fixes ou adaptatives. Cependant, la plupart de ces stratégies reposent sur une borne connue (en temps ou en tours). Dans les applications quotidiennes, par exemple lorsque nous cherchons à acquérir un bien de consommation non critique, il est commun que le protocole ne propose aucune borne, et que chaque agent ignore le moment où il coupera court à la négociation. 
	
	Dans cet article, nous étudions MoCaNA ({\bf Mo}nte-{\bf Ca}rlo {\bf N}egotiating {\bf A}gent), un agent de négociation automatique conçu pour le protocole de \textsl{bargaining} (ou marchandage). Dans ce protocole, deux agents s'échangent des offres de manière à trouver un accord. La négociation s'interrompt lorsqu'un des deux adversaires accepte l'offre qui lui est faite par son adversaire ou qu'il la rejette. MoCaNA a pour particularité de ne faire aucune présupposition sur: 1) la linéarité des préférences, 2) le caractère continu ou discret de l'espace de négociation ni 3) l'imposition d'une borne par le protocole. Il exploite pour cela des connaissances issues des domaines du \textsl{General Game Playing} et de \textsl{l'Apprentissage Automatique}. Afin de décider de la valeur d'une offre, notre agent s'appuie sur la recherche arborescente de Monte-Carlo (MCTS), une heuristique qui a été utilisée avec succès dans de nombreux jeux comme le go, qu'il combine avec de la modélisation d'adversaires pour plus d'efficacité.
	
	Le choix de cette heuristique est guidé par deux éléments. D'abord, MCTS s'est montré efficace dans des jeux très différents, ce qui lui a valu de nombreux succès en \textsl{General Game Playing}. Il est aussi adapté aux jeux à haut facteur de branchement. Il constitue donc une bonne alternative pour les domaines de négociation complexes dont le facteur de branchement à chaque proposition est l'ensemble des propositions possibles.
	
	Cet article est oorganisé comme suit: la \cref{relatedworks} introduit les travaux de la littérature traitant de négociation automatique et des méthodes de Monte-Carlo. Nous proposons ensuite une modélisation de la négociation comme un jeu dans la \cref{negogame}. La \cref{secmocana} présente les fonctionnalités de MoCaNA. Nous présentons les résultats de la confrontation de MoCaNA aux finalistes de l'ANAC 2014 et au RandomWalker dans la \cref{xp}. Nous résumons ces contributions et donnons des pistes d'amélioration de notre travail dans la \cref{conclusion}.
	
	\section{Travaux connexes}
	\label{relatedworks}
	MoCaNA se trouve au croisement des domaines de la négociation automatique et des techniques de Monte-Carlo appliquées aux jeux. Nous donnons dans cette section une introduction à ces deux domaines.
	
	\subsection{Négociation automatique}
	Les architectures des agents de négociation automatique impliquent trois fonctionnalités \cite{Baarslag2016Exploring}:
	\textbf{la stratégie d'offres} définit les offres que l'agent fait à son adversaire;
	\textbf{la stratégie d'acceptation} définit si l'agent accepte la proposition qui lui a été faite ou s'il fait une contre-offre;
	\textbf{la modélisation d'adversaires} permet de modéliser certains aspects de l'adversaire, comme sa stratégie d'offre, son utilité ou sa stratégie d'acceptation et est utilisé pour améliorer l'efficacité de la stratégie d'offres de l'agent et parfois celle de sa stratégie d'acceptation.
	
	\subsubsection{Stratégie d'offres}
	Les stratégies d'offres utilisées dans le cadre de négociations complexes s'appuient sur deux familles de techniques, que nous présentons par la suite: les heuristiques couplées au domaine et les algorithmes génétiques.
	
	Les heuristiques s'appuient généralement sur des tactiques \cite{Faratin1998Negotiation}, qui consistent à faire des concessions selon un élément de la négociation comme le temps écoulé, la rareté de la denrée négociée, ou les concessions faites par l'adversaire. Il est possible d'adapter ces tactiques ou de sélectionner une offre ayant une utilité gagnant-gagnant s'il en existe une. La majorité de ces stratégies \cite{Faratin1998Negotiation} reposent sur la connaissance de la borne, ce qui les rend inapplicables dans notre contexte.
	
	Les algorithmes génétiques \cite{Jonge2016} sélectionnent les offres qui maximisent une fonction objectif déterminée, les croisent et les modifient de manière à les améliorer. Ces stratégies génèrent un certain nombre d'offres parmi lesquelles l'agent doit choisir celle qui sera envoyée à l'adversaire. Elles ont obtenu de bons résultats, sans toutefois vaincre les agents utilisant des heuristiques.
	
	\subsubsection{Stratégies d'acceptation}
	
	On distingue deux principales catégories de stratégies d'acceptation \cite{Baarslag2013}: les stratégies myopes et les stratégies optimales. Les stratégies myopes reposent sur les offres de l’adversaire et celles de l'agent. Il peut s'agir d'accepter toute offre dépassant un certain seuil, toute offre meilleure que la dernière faite par l'agent lui-même ou que la prochaine générée par stratégie d'offre. Il est aussi possible de combiner ces stratégies. Les stratégies optimales \cite{Baarslag2013} exploitent un modèle de la stratégie adverse afin de déterminer la probabilité d'obtenir une meilleure offre.
	
	\subsubsection{Modélisation de l'adversaire}
	La majorité des méthodes de modélisation d'adversaire utilisées en négociation automatique ont été analysées par Baarslag et ses collègues \cite{Baarslag2015Learning}. Elles permettent de modéliser la stratégie d'offre, la stratégie d'acceptation, l'utilité de l'adversaire, la borne qu'il s'est fixé, et son prix de réserve le cas échéant.
	
	Deux familles de techniques sont adaptées à la modélisation de \textbf{stratégies d'offres} adaptatives et non bornées : les réseaux de neurones et les techniques reposant sur l'analyse de séries temporelles. Les approches basées sur l'analyse de séries temporelles sont diverses. Parmi elles, la régression de processus gaussien est stochastique. Elle s'appuie sur les mouvements précédents pour prédire une densité de probabilité pour le tour suivant. Introduite par Rasmussen et Williams \cite{Rasmussen2006Gaussian}, elle a été utilisée avec succès par Williams \textsl{et al.} \cite{Williams2011Using}.
	L'aspect stochastique de cette méthode permet de générer des propositions différentes à chaque étape de simulation de \textsl{Monte Carlo Tree Search} (MCTS), selon la gaussienne prédite par la régression. C'est cette méthode que nous adoptons.
	
	L'\textbf{utilité} d'un adversaire est généralement considérée comme la somme pondérée de fonctions pour chaque attribut à valeurs dans $[0,1]$ \cite{Baarslag2015Learning}. Deux familles de méthodes sont utilisées. La première s'appuie sur la fréquence d'apparition de chaque valeur parmi les propositions de l'adversaire. Elle suppose que les valeurs proposées le plus régulièrement par l'adversaire sont celles qu'il préfère, et que les attributs variant le plus souvent sont ceux qui ont le moins d'importance pour lui. Ces approches ne sont cependant valables que dans le cas où le domaine de négociation est discret. En effet, l'extension au cas continu requiert la définition d'une fonction de distance dépendant du domaine.
	La seconde famille de méthodes repose sur l'apprentissage bayésien. Elle est adaptée au cas continu. La méthode présentée par Hindriks et Tykhonov \cite{Hindriks2008Opponent} repose sur la génération d'hypothèses. Chaque hypothèse est constituée d'un ordonnancement des attributs, et d'une fonction d'utilité simple (linéaire ou linéaire par morceaux) pour chacun d'entre eux. Le poids associé à chaque attribut est calculé en fonction de l'ordre qui lui est attribué. La fonction d'utilité estimée est la somme des hypothèses pondérée par leur probabilité.
	
	La \textbf{stratégie d'acceptation} d'un adversaire peut être apprise de deux manières. On peut d'abord faire la supposition que l'adversaire acceptera une offre sous certaines conditions dépendant de sa stratégie d'offre et/ou de sa fonction d'utilité \cite{Baarslag2015Learning}. Dans ce cas, il est possible de la déduire des modèles ci-dessus sans faire de calcul supplémentaire. Dans le cas où l'agent ne modélise pas les éléments précédemment décrits, il est aussi possible d'utiliser des réseaux de neurones \cite{Fang2008Opponent}. Cela demande cependant un calcul supplémentaire, potentiellement coûteux.
	
	\subsection{Méthodes de Monte Carlo}
	
	Les méthodes de Monte-Carlo sont utilisées comme heuristiques pour les jeux. Elles attribuent à chaque mouvement possible un score calculé au terme d'une ou de plusieurs simulations. Kocsis et Szepevsv\'ari \cite{Kocsis2006Bandit} proposent une méthode hybridant la construction d'un arbre de jeu, méthode qui a fait ses preuves, et les méthodes de Monte-Carlo. Chaque nœud de l'arbre de jeu garde en mémoire les scores obtenus lors des simulations où il a été joué, ainsi que le nombre de simulations faites dans la branche partant du nœud. Pendant l'exploration de cet arbre, les branches privilégiées sont celles qui ont été peu explorées et celles qui ont le plus haut score dans les simulations.
	Cette méthode, nommée recherche arborescente de Monte-Carlo (\textsl{MCTS}\footnote{\textsl{Monte Carlo Tree Search}}) s'est vue améliorée au moyen de nombreuses extensions \cite{Browne2012survey}. 
	
	MCTS se décompose en quatre parties. (1) La première étape consiste à parcourir l'arbre déjà constitué selon une stratégie prédéfinie. À chaque nœud, on décide s'il faut sélectionner une branche de niveau inférieur ou en explorer une nouvelle. (2) Dans le second cas, l'arbre se voit pourvu d'un nouveau nœud, généré selon une stratégie d'expansion. (3) Ensuite, on simule le jeu jusqu'à un état final. (4) On calcule les scores et on les rétropropage sur les nœuds de l'arbre ayant été explorés.
	
	Les méthodes de Monte-Carlo ont gagné en popularité suite à leur succès dans les jeux à fort facteur de branchement, notamment le Go. En particulier, AlphaGo \cite{Silver2016}, qui a vaincu l'un des meilleurs joueurs au monde, utilise les méthodes de Monte-Carlo couplées à de l'apprentissage profond. À notre connaissance, seuls de~Jonge et Zhang \cite{Jonge2017Automated} ont utilisé MCTS dans le cadre de la négociation automatique. Ils se sont concentrés sur des domaines de négociations petits, et où la fonction d'utilité adverse est facilement inférable, comme la répartition d'un dollar entre deux joueurs. Nous nous concentrons sur des domaines plus complexes, où la fonction d'utilité adverse ne peut être facilement inférée et où l'espace de négociation peut être grand, voire infini.

	\section{Jeu et négociation}
	
	\label{negogame}
	La recherche arborescente de Monte-Carlo est une méthode appliquée aux jeux extensifs. Dans cette section, nous montrons comment il est possible d'assimiler la négociation à un tel jeu. Nous commençons par associer chaque aspect de la négociation à un élément d'un jeu. Nous décrivons ensuite les particularités de la négociation, qui empêchent l'utilisation des stratégies les plus répandues dans MCTS. Nous concluons cette section en donnant d'autres formalismes possibles pour la négociation et en expliquant les relations qu'ils entretiennent avec celui que nous avons choisi.
	

	\subsection{Notre formalisme}
	Un jeu extensif \cite{Osborne1994} est composé d'un ensemble de joueurs, de la description des historiques de jeu possibles, d'une fonction indiquant le tour de chaque joueur et d'un profil de préférence. En s'appuyant sur cette définition, il est possible de définir un \textsl{bargaining} $\mathcal{B}$ comme un jeu sous forme extensive.
	\begin{definition}[\textsl{Bargaining}]
		\label{ch4-bargaining}
		Un \textsl{bargaining} est un triplet $\mathcal{B} = \left(H,A,\left(u_i\right)_{i\in\llbracket 1,2\rrbracket}\right)$ vérifiant:
		\begin{enumerate}
			\item \textbf{joueurs:} il y a deux joueurs: un acheteur (joueur $1$) et un vendeur (joueur $2$),
			\item \textbf{historique: } l'historique de la négociation est composé des messages que les agents s'envoient: les propositions faites par les agents, et les \texttt{accept} et \texttt{reject}. Les historiques terminaux sont les historiques infinis et les historiques se terminant par \texttt{accept} ou par \texttt{reject}. Chaque message est composé d'une paire $(\alpha,c)$ où $\alpha$ est l'acte de langage (\textsl{performatif}) du message et, $c$ est le contenu du message, \textsl{i.e.} une liste de couples $(k,v)$ où $k$ est la clé d'un attribut du domaine de négociation et $v$ la valeur correspondante. On sépare tout historique en deux, chacun correspondant aux actions d'un joueur: $h_i = (\alpha,c)_i$
			\item \textbf{tour de jeu:} la fonction tour fonctionne selon la parité. On suppose que c'est l'acheteur (le joueur 1) qui commence. Ainsi, $\forall h\in H, joueur(h)=2-(|h| \mod 2)$ où $|h|$ est la taille de $h$,
			\item \textbf{préférence: }l'ordre sur les situations terminales de chaque joueur est induit par une fonction d'utilité $u_i$ qui associe son utilité pour chaque historique terminal. Cette fonction associe une utilité à chaque accord possible trouvé, ainsi qu'au cas de rejet et au cas d'historique infini\footnote{Notons que ces deux dernières valeurs peuvent être différentes, par exemple si l'agent considère qu'il alloue des ressources à la négociation.}.
		\end{enumerate}
	\end{definition}
	
	La négociation n'est pas un jeu combinatoire classique, comme peuvent l'être les échecs ou le go. La première différence entre ces jeux et la négociation tient au fait que cette dernière est un jeu à somme non nulle. Les agents cherchent à trouver un accord mutuel qui soit profitable à chacun. Cela est particulièrement vrai dans des domaines complexes, ayant de nombreux attributs, où une solution peut être bien meilleure que la solution par défaut où chaque agent reçoit son utilité de réserve, \textsl{i.e.} quand les agents ne trouvent pas d'accord.
	
	Ensuite, la négociation est un jeu à information incomplète, c'est à dire que le type des joueurs, ici leur profil de préférence, est inconnu et fait même généralement l'objet d'une modélisation. Ces deux particularités rendent inutilisable l'\textsl{Upper Confidence Tree \cite{Kocsis2006Bandit}} utilisé pour les jeux combinatoires comme le go car il est fait pour les jeux combinatoires. Enfin, le domaine de la négociation est très particulier. Il peut être catégoriel (\textsl{e.g.} couleur) mais également numérique voire continu. Cela a des conséquences sur l'exploration de l'arbre, en particulier sur le critère décidant de l'expansion d'un nouveau nœud. Malgré ces difficultés d'adaptation, les succès qu'a remporté MCTS dans les jeux complexes semblent indiquer qu'il pourrait s'avérer être une bonne stratégie pour la négociation dans un contexte complexe.
	
	\subsection{Autres formalismes}
	
	Bien que nous utilisions le formalisme des jeux extensifs pour modéliser la négociation, d'autres modélisations sont possibles, notamment en utilisant les jeux bayésiens et les jeux stochastiques.
	
	Les \textbf{jeux à information incomplète} peuvent être modélisés en utilisant le formalisme des jeux bayésiens \cite{Reeves2004Computing}. Ces jeux sont traditionnellement utilisés pour modéliser les jeux à information incomplète, mais supposent en général la connaissance d'une probabilité \textsl{a priori} sur les types possibles de l'adversaire. Notons que dans le cadre que nous nous sommes fixés, il n'existe pas de telle distribution \textsl{a priori}. La prise en compte de l'information révélée par l'adversaire au cours de la négociation, qui permet d'établir des probabilités sur le profil de préférence, se fait au cours de la négociation en utilisant la modélisation de l'adversaire.
	
	Un autre formalisme possible est la modélisation par un \textbf{jeu stochastique} \cite{Jaskiewicz2016}.
	Les jeux stochastiques sont une généralisation des processus markoviens. Le \textsl{bargaining} peut être vu comme un processus de décision markovien (MDP), où chaque action de l'agent génère une réaction de son adversaire, amenant une transition vers un autre état du jeu. Dans ce nouvel état, les actions possibles restent les mêmes, mais l'utilité induite par l'acceptation de la proposition de l'adversaire change. L'exploitation de ce formalisme se fait par une exploration des transitions possibles et une pondération par leur probabilité. MoCaNA cherche à évaluer ces probabilités en utilisant la modélisation de l'adversaire et estime l'utilité attendue pour les différentes transitions possibles, \textsl{i.e.} les choix de l'adversaire, au moyen de MCTS. Notons que cette méthode s'est montrée particulièrement efficace pour la planification dans les MDP, comme le montre notamment \cite{Kocsis2006Bandit}.

	\section{MoCaNA}
\begin{figure}[!htbp]
		\centering
		\begin{tikzpicture}[scale=.7, every node/.style={scale=.7}]
		\node at (4,3.6) [draw,  anchor=south east, minimum height=4ex] (selection) {Sélection};
		\node at (6.5,3.6) [draw, anchor=south east, minimum height=4ex] (expansion) {Expansion};
		\node at (9,3.6) [draw, anchor=south east, minimum height=4ex] (simulation) {Simulation};
		\node at (12.5,3.6) [draw, anchor=south east, minimum height=4ex] (backprop) {Rétropropagation};
		
		\node at (7.375,5) {Stratégie d'offres};
		
		\draw (2,5.5) -- (12.75,5.5) -- (12.75,3.5) -- (2,3.5) -- cycle;
		
		\node at (5,2.4) [draw, text width=5.5cm, text centered, anchor=north] (strategy) {Strategie (Régression de processus gaussien)};
		\node at (10.5,2.4) [draw, text width=4cm, text centered, anchor=north] (utility) {Utilité (Apprentissage Bayesien)};
		
		\node at (7.375,1) {Modélisation d'adversaire};
		
		\draw (2,2.5) -- (12.75,2.5) -- (12.75,.5) -- (2,.5) -- cycle;
		
		\node at (13.5,3) [text width=5cm, text centered, rotate=-90] {Strategie d'acceptation};
		\draw (13,5.5) -- (13,.5) -- (14, .5) -- (14,5.5) -- cycle;
		
		\draw [->, >=latex] (simulation) -- (strategy);
		\draw [->, >=latex] (simulation) -- (utility);
		\draw [->, >=latex] (backprop) -- (utility);
		\draw [->, >=latex] (13,3) -- (12.75,3.5);
		
		\end{tikzpicture}
		\caption{Interaction entre les modules de MoCaNA}
		\label{interaction}
	\end{figure}
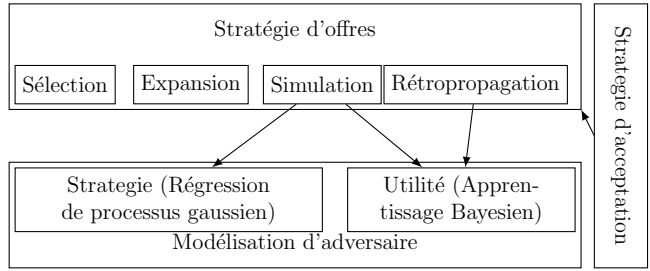

	Comme nous l'avons expliqué dans la section précédente, la négociation est un jeu particulier. Il est donc nécessaire d'ajuster les heuristiques pour les jeux à ces particularités. Dans cette section, nous présentons notre agent de négociation automatique exploitant MCTS. L'architecture générale des différents modules de notre agent est présentée dans la \Cref{interaction}. La stratégie d'offre implémente MCTS et utilise le module de modélisation d’adversaire, qui comporte deux sous-modules: l'un pour l’utilité de l’adversaire, l’autre pour sa stratégie d’offre. Le dernier module, celui de stratégie d'acceptation, effectue une comparaison entre la proposition de l'adversaire et la proposition générée par la stratégie d'offre. Les différents sous-modules de l'agent ainsi que les interactions entre eux sont décrits dans la suite de cette section.
	\label{secmocana}
	
	\subsection{Modélisation d'adversaire}
	\label{ch5-2-opponent-modeling}
	
	Afin de d'améliorer l'efficacité de MCTS, il est nécessaire de disposer d'un modèle de sa stratégie d'offre ainsi qu'un modèle de son profil de préférence.
	
	\subsubsection{Modélisation de la stratégie d'offre}
	\label{ch5-2-Learn}
	
	Le but est de prédire une future proposition faites par l'adversaire au tour $x_*$ de la négociation. Nous utilisons la régression de processus gaussiens \cite{Rasmussen2006Gaussian} qui permet de générer une gaussienne centrée en la valeur prédite par l'algorithme et dont l'écart type représente l'incertitude du modèle.
	
	On commence par calculer la matrice de covariance $K$, qui représente la proximité entre les tours $(x_i)_{i\in\llbracket 1, n\rrbracket}$ de la séquence, en nous reposant sur une fonction de covariance, aussi appelée \textsl{kernel} $k$.
	
	Soit \begin{equation}\def\arraystretch{0.7}
		K= \left( \begin{array}{ccc}
			k(x_1, x_1) & \dots & k(x_1, x_n) \\
			\vdots &  & \vdots \\
			k(x_n, x_1) & \dots & k(x_n, x_n) \end{array} \right)
	\end{equation}
	
	on calcule ensuite la distance entre le tour dont on veut estimer la proposition $x_*$ et les tours précédents:
	
	\begin{equation}K_* = \left(k(x_*, x_1), \dots, k(x_*, x_n)\right)\end{equation}
	
	La régression de processus gaussien fait la supposition que chacune de ces valeurs est la composante d'une gaussienne multivariée. On en déduit:
	\begin{equation}
		\overline{y_*} = K_*K^{-1}\mathbf{y}
	\end{equation}
	\begin{equation}
		\sigma^2_* = \mathrm{Var}(y_*) = K_{**}-K_*K^{-1}K_*^\top
	\end{equation}
	où $K_{**} =k(x_*, x_*)$. Nous identifions $y_*$ à une variable aléatoire gaussienne de moyenne $\overline{y_*}$ et d'écart type $\sigma_*$.
	
	Une des composantes les plus importantes de la régression de processus gaussien est la détermination du \textsl{kernel}. Les plus communs sont les fonctions à base radiale, les fonctions \textsl{rational quadratic}, le \textsl{kernel} de Matérn et le \textsl{kernel exponential sine squared}.  Ces kernels servent à déterminer la distance entre deux tours de négociation. Nous avons testé ces quatre \textsl{kernels} sur plusieurs négociations entre agents. La \Cref{kernel-table} donne les résultats de la régression de processus gaussien pour chacun de ces \textsl{kernels}. Nous avons fait ces tests dans le contexte de l'ANAC 2014, où les préférences ne sont pas linéaires, et avec les finalistes de l'ANAC. Nous avons généré aléatoirement 25 sessions, et avons modélisé les deux agents, obtenant ainsi 50 modélisations au total. Chaque offre de chaque série est prédite en fonction des précédentes et sert à prédire les suivantes. La table montre la distance euclidienne moyenne entre les propositions faites et le résultat de l'apprentissage. Plus la valeur est basse, plus la prédiction est proche de la véritable série.
	
	\begin{table}[!ht]
		\setlength{\tabcolsep}{12pt}	
		\centering
		\begin{tabular}{r l}
			\textsl{Kernel} & Dist. moyenne\\\hline
			Fonction à base radiale & 43.288\\
			\textbf{\textsl{Rational quadratic}} & \textbf{17.766}\\
			Matérn & 43.228\\
			\textsl{Exp. sine squared} & 22.292
		\end{tabular}
		\caption{Distance moyenne entre les propositions et les prédictions de la régression de processus gaussien selon le \textsl{kernel}}
		\label{kernel-table}
	\end{table}
	
	Le \textsl{kernel} donnant le meilleur résultat est le \textsl{Rational quadratic}. C'est donc celui-ci que notre agent utilise.
	
	\subsubsection{Modèle du profil de préférence}
	\label{ch5-2-opp-modeling-utility}
	L'apprentissage bayésien décrit dans \cite{Hindriks2008Opponent} suppose seulement que l'agent fait une concession à un rythme constant, une supposition faible par rapport aux autres méthodes.
	
	On approxime l'utilité de l'adversaire par une somme pondérée de fonctions triangulaires. Une fonction $t$ de $[a,b]\subset\mathds{R}$ dans $[0,1]$ est dite triangulaire si et seulement si:\begin{itemize}
		\item $t$ est affine, et $t(a)=0$ et $t(b)=1$ ou $t(a)=1$ et $t(b)=0$ ou
		\item il existe $c$ dans $[a,b]$ tel que $t$ est affine sur $[a,c]$ et sur $[c,b]$, $t(a)=0$, $t(b)=0$, $t(c)=1$.
	\end{itemize}
	La méthode se divise en deux étapes. D'abord, un certain nombre d'hypothèses sont générées. Ces hypothèses sont composées d'une somme pondérée de fonctions triangulaires. Chaque attribut se voit attribuer un poids et une fonction triangulaire.
	
	L'utilité estimée d'un adversaire est la somme de ces hypothèses pondérée par la probabilité de chacune. Cette méthode a pour avantage de ne pas faire de supposition sur la stratégie de l'adversaire, mais elle suppose qu'il fait des concessions à un rythme globalement linéaire.
	
	\subsubsection{Modèle de stratégie d'acceptation}
	La stratégie d'acceptation des agents est modélisée de manière très simple: on considère qu'un agent accepte la dernière proposition faite par son adversaire si et seulement si son utilité est meilleure que l'utilité générée par la stratégie d'offre. Cette modélisation, évoquée par \cite{Baarslag2015Learning}, est peu coûteuse et permet de ne pas alourdir les calculs déjà nombreux de l'agent.
	
	\subsection{Stratégie d'offres basée sur MCTS}
	Les méthodes de Monte-Carlo sont très adaptatives, et ont obtenu de bons résultats dans différents jeux, y compris des jeux à grand facteur de branchement. Dans cette section, nous introduisons une stratégie d'offres basée sur MCTS.
	
	\subsubsection{MCTS sans élagage}
	Comme nous l'avons expliqué, MCTS est un algorithme général, qui repose sur un certain nombre de stratégies. À chaque fois que l'agent doit prendre une décision, il génère un nouvel arbre et l'explore en utilisant MCTS. L'implémentation la plus commune de MCTS est l'\textsl{Upper Confidence Tree} (UCT). Cette méthode a notamment donné de très bons résultats pour le Go. Néanmoins, lors de la phase de sélection, cette implémentation étend un nouveau nœud dès lors que tous les enfants d'un arbre n'ont pas été étendus. Or, dans le cas d'un domaine de négociation contenant un attribut infini, le nombre de fils possibles pour un nœud est infini, puisqu'il peut prendre toutes les valeurs possibles pour cet attribut. L'algorithme étendra donc indéfiniment de nouveaux nœuds sans jamais explorer l'arbre en profondeur. Si l'arbre est très large, si le nombre de nœuds possibles est plus grand que nombre de simulations, la même situation apparaîtra.
	
		Dans le contexte de la négociation, il est donc nécessaire de définir une implémentation de MCTS différente:
	\begin{description}
		\item[Sélection] Pour cette étape, nous utilisons le \emph{progressive widening}. Le critère du progressive widening peut s'exprimer ainsi: un nouveau nœud est étendu si et seulement si:
		\begin{equation}
			n_p^\alpha\geq n_c
			\label{prog_wid_equation1}
		\end{equation}
		où $n_p$ est le nombre de fois où le parent a été simulé, $n_c$ est le nombre de fils qu'il a, et $\alpha$ est un paramètre du modèle. S'il n'y a pas d'expansion, nous sélectionnons le nœud $i$ qui maximise:\begin{equation}
			W_i = \dfrac{s_i}{n_i+1} + C\times n^\alpha\sqrt{\frac{\ln(n)}{n_i+1}}
			\label{prog_wid_equation2}
		\end{equation}
		où $n$ est le nombre total de simulations faites jusqu'alors, $s_i$ est le score du nœud $i$ et $C$ est également un paramètre du modèle.
		\item[Expansion] La valeur d'un nœud étendu est quant à elle choisie de manière aléatoire, avec une distribution uniforme sur le domaine de négociation. 
		\item[Simulation] Nous utilisons le modèle de stratégie de l'adversaire, afin de rendre les simulations plus pertinentes. Nous utilisons la régression de processus gaussien et l'apprentissage bayésien.
		\item[Rétropropagation] L'étape de rétropropagation utilise aussi le modèle de l'utilité de l'adversaire, et répercute sur les nœuds l'utilité de l'agent et l'utilité de son adversaire selon la modélisation d'utilité.
	\end{description}
	
	Enfin, MCTS fait des simulations et calcule l'utilité attendue de toute la négociation. Il a tendance à sous-estimer la valeur de la proposition qu'il ressort, et la probabilité que l'adversaire l'accepte. Nous renvoyons donc l'offre $b^*$ parmi les fils de la racine maximisant:
	\begin{equation}
		b^*=\argmax_{b \in racine.fils}\left(\frac{utility(b) + score(b)}2\right)
	\end{equation}
	
	\subsubsection{Élagage}
	\label{ch5-mcts-pruning}
	Afin de n'explorer que les nœuds donnant une utilité importante à l'agent, il est possible d'utiliser des connaissances sur le jeu pour élaguer certaines parties de l'arbre et ne pas les développer.	Nous considérons deux cas, un élagage dit \og fixe\fg{} et un élagage dit \og variable \fg.
	
	Dans le cas de l'\textbf{élagage fixe}, nous fixons à notre agent une utilité minimale attendue, \textsl{i.e.} une utilité de réserve. Au sein de la simulation, lorsque MoCaNA produit une proposition, l'utilité de cette proposition est calculée en utilisant le profil d'utilité de MoCaNA. Si cette utilité est inférieure à la valeur seuil, la branche est supprimée.
	
	Lorsque l'agent n'est pas capable de déterminer un prix de réserve, il est possible de procéder à un \textbf{élagage variable} dépendant des propositions de l'adversaire. Dans ce cas, l'agent garde en mémoire la proposition de l'adversaire ayant pour lui la meilleure utilité. Toute proposition générée par MoCaNA dans une simulation ayant une utilité inférieure est élaguée. Avec cette politique, l'agent ne fait que des propositions meilleures que celles que son adversaire a faites.
	
	\subsection{Architecture logicielle}
\begin{figure}
		\centering
		\begin{tikzpicture}[->, ultra thick, >=latex, every node/.style={scale=.8},scale=.8]
			\node at (6,0) (genius){\textcolor{lblue}{\fontsize{50pt}{5pt}\faPlug}};
			\node at (3,-3) (mcts){\textcolor{lred}{\fontsize{50pt}{5pt}\faGears}};
			\node at (6,-3) (agent){\textcolor{lgreen}{\includegraphics[height=50pt]{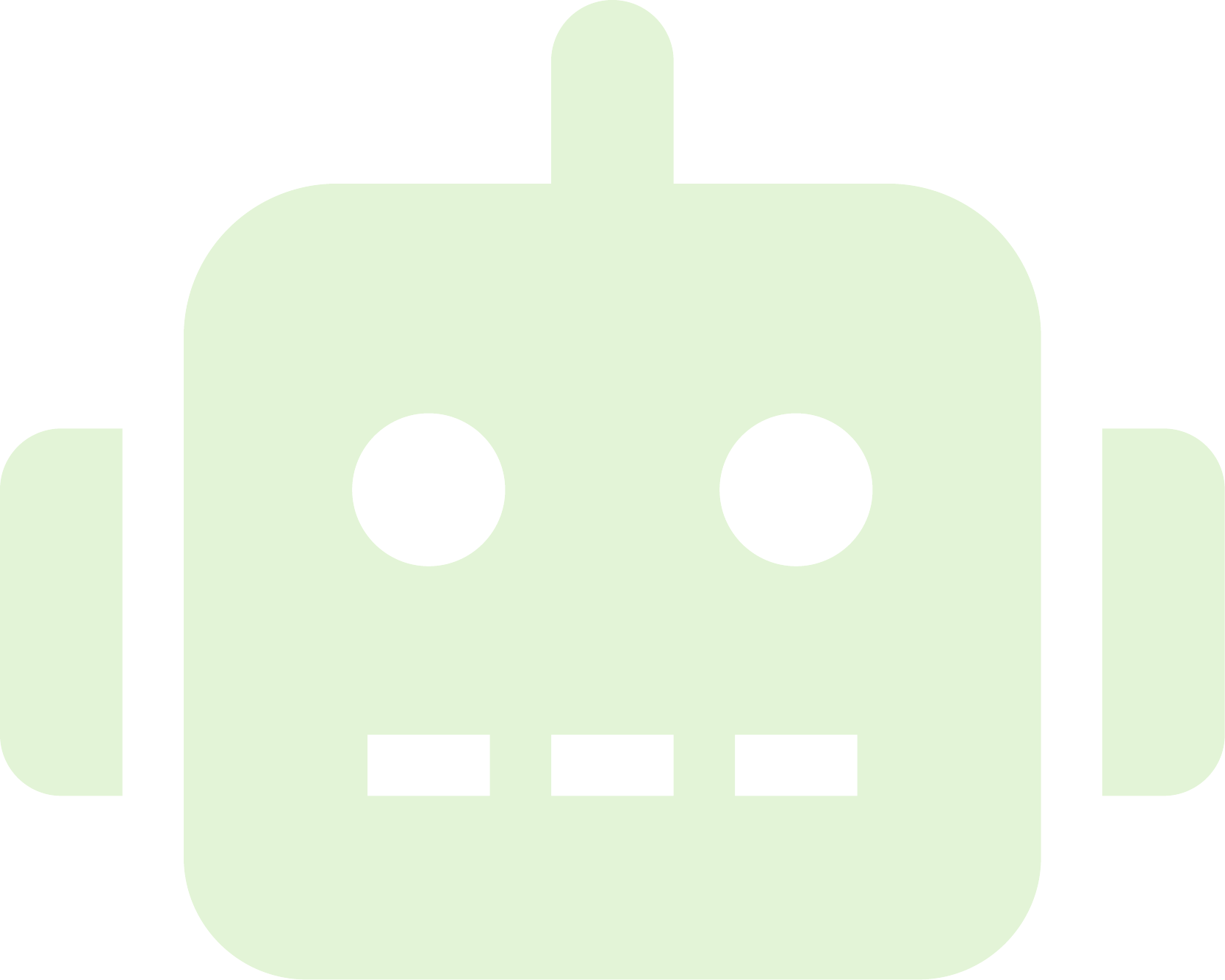}}};
			\node at (0,0) (strategy){\textcolor{lviolet}{\fontsize{50pt}{5pt}\faLineChart}};
			\node at (0,-3) (utility){\textcolor{lorange}{\fontsize{50pt}{5pt}\faBarChart}};
			\node at (3,0) (data){\textcolor{lgray}{\fontsize{50pt}{5pt}\faDatabase}};
			
			\node at (6,0) [text width=3cm, text centered]{Connecteur \textsc{Genius}};
			\node at (3,-3) [text width=2.5cm, text centered]{Stratégie d'offre (MCTS)};
			\node at (6,-3) [text width=3cm, text centered]{Agent};
			\node at (0,0) [text width=3cm, text centered]{Modélisation Stratégie};
			\node at (0,-3) [text width=3cm, text centered]{Modélisation Utilité};
			\node at (3,0) [text width=3cm, text centered]{Données};

			\draw [gray](7,-3.5) -- (8,-3.5) node [midway, below] {\faEnvelope};
			\draw [gray](8,-2.5) -- (7,-2.5) node [midway, above] {\faEnvelope};
			\draw [gray](agent) -- (data) node [midway, below left] {\faEnvelope};
			\draw [gray](mcts) -- (strategy);
			\draw [gray](mcts) -- (utility);
			\draw [gray](agent) -- (mcts);
			\draw [gray](agent) -- (genius);
			\draw [gray](strategy) -- (data);
			\draw [gray](utility) -- (data);
		\end{tikzpicture}
		\caption{Architecture logicielle des modules de MoCaNA}
		\label{architecture}
	\end{figure}
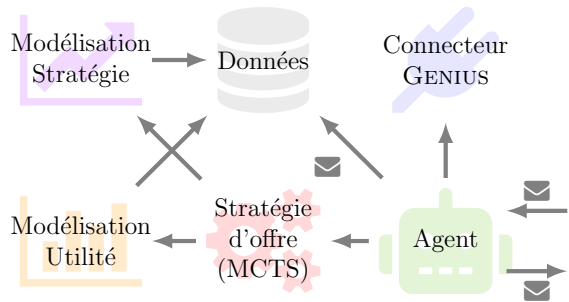

	Notre agent a été développé en java. Nous l'avons conçu de façon à ce qu'il soit modulaire, et à ce qu'il soit découplé du \textsl{framework} sur lequel les expérimentations ont été réalisées. Nous avons pour cela construit un module permettant à notre agent de s'interfacer avec \textsc{Genius}. L'agent est organisé sous la forme d'interfaces implémentées par les classes de chaque module, et qui sont traduites pour \textsc{Genius} le cas échéant. On retrouve dans l'architecture de MoCaNA les différents éléments décrits ci-dessus: la stratégie d'offre incluant MCTS et les concepts allant avec (\textsl{e.g.} notions d'arbre, de nœud), un module de modélisation de la stratégie adverse et un modèle de l'utilité. Nous n'avons pas implémenté de module pour la modélisation de la stratégie d'acceptation. Cette architecture est représentée dans la \Cref{architecture}. Les \faEnvelope{} représentent les messages envoyés et reçus par l'agent.
	
	Les simulations de MCTS étant très gourmandes en terme de puissance de calcul, nous les avons parallélisé. L'aspect stochastique de notre modélisation de stratégie d'offre assure que deus simulations lancées du même point n'auront pas le même résultat, tout en privilégiant les valeurs les plus probables. La régression de processus gaussien, qui repose sur le calcul matriciel, a été développée en utilisant la librairie JAMA. Les paramètres du kernel de cette méthodes sont quant à eux optimisés en utilisant la librairie Apache Commons Math.
	
	\section{Expérimentations}
	Pour évaluer notre agent, nous utilisons le \textsl{framework} \textsc{Genius} \cite{Lin2014}, qui permet de créer des sessions de négociation et des tournois. Nous confrontons MoCaNA au RandomWalker ainsi qu'aux agents de l'ANAC 2014.
	\label{xp}
	\subsection{Protocole expérimental}
	\textsc{Genius} permet de négocier sur des attributs entiers ou discrets, mais pas encore continus. Nous utilisons le domaine de négociation de l'ANAC~2014 décrit par \cite{Fukuta2016}. Les compétitions de 2015 et de 2016 se sont concentrées respectivement sur la négociation multilatérale, et sur les \textsl{smart grids}. À notre connaissance, les agents de négociation de l'ANAC 2017 n'ont pas encore été mis à disposition. Les attributs sont entiers, et varient de 0 à 10. Plusieurs variantes ont été proposées allant de 10 attributs à 50 attributs. Nous nous focalisons sur le contexte à 10 attributs. Le contexte utilisé n'a aucun taux de réduction. L'utilité est une somme pondérée de fonctions non linéaires sur chaque attribut. L'utilité de réserve des agents (dans le cas où ils ne trouvent pas d'accord) est 0, c'est à dire la valeur minimale.
	
	Nous fixons en outre la borne de chaque session de négociation à 1 heure. Après ce temps, la négociation s'interrompt et les agents obtiennent leur utilité de réserve, c'est-à-dire 0. Nous calibrons notre modèle de manière empirique. Lorsque notre agent peut choisir parmi 200 propositions, il s'en trouve généralement une qui procure à la fois une haute utilité pour lui et son adversaire. Afin que l'arbre soit également exploré en profondeur, nous lui laissons assez de temps pour qu'il génère en moyenne 50.000 simulations. Ainsi, chaque pour faire une proposition, notre agent prend 3 minutes. Nous obtenons $\alpha=0.489$.
	
	\subsection{Résultats}
	\label{ch5-3-resultats}
	
	Nous distinguons deux contextes de négociation: un contexte où la borne est présente et connue des adversaires, face aux agents de l'ANAC 2014, et un où elle est assez loin pour être considérée comme inexistante, face au \textsl{RandomWalker}.
	
	\subsubsection{ANAC~2014}
	
	\begin{sidewaystable} 
		\setlength{\tabcolsep}{7pt}
		\centering
		{\small
			\begin{tabular}{r | c c c}
				Adversaire & Score adversaire & Score MoCaNA & Taux d'accord\\\hline
				AgentM & 0.88 $(\pm0.07)$/0.70$(\pm0.08)$/0.77$(\pm0.11)$ & 0.66 $(\pm0.08)$/0.85 $(\pm0.02)$/0.66$(\pm0.08)$ & 0.35/0.15/0.10\\
				DoNA & 1$(\pm0.0)$/N.A/0.86$(\pm0.01)$ & 0.57 $(\pm0.0)$/N.A/0.55$(\pm0.06)$ & 0.05/0/0.10\\
				Gangster & 0.81 $(\pm0.13)$/0.53$(\pm 0.19)$/0.56$(\pm0.22)$ & 0.4 $(\pm0.14)$/0.85$(\pm 0.05)$/0.70$(\pm0.16)$ & 0.45/0.55/0.55\\
				Whale & 0.77 $(\pm0.16)$/0.71$(\pm0.06)$/0.64$(\pm0.09)$ & 0.70 $(\pm0.11)$/0.83$(\pm0.02)$/0.77$(\pm0.10)$ & 0.45/0.45/0.25\\
				Group2 & 0.68 $(\pm0.21)$/0.53$(\pm0.06)$/0.53$(\pm0.08)$ & 0.82 $(\pm0.19)$/0.87$(\pm0.04)$/0.79$(\pm0.10)$ & 0.45/0.45/0.45\\
				kGAgent & 0.98 $(\pm0.085)$/N.A/1$(\pm0)$ & 0.55 $(\pm0.05)$/N.A/0.49$(\pm0)$ & 0.20/0/0.05\\
				AgentYk & 0.81 $(\pm0.17)$/0.71$(\pm0.07)$/0.77$(\pm0.13)$ & 0.752 $(\pm0.15)$/0.84$(\pm0.03)$/0.64$(\pm0.11)$ & 0.15/0.10/0.15\\
				BraveCat & 0.62$(\pm0.17)$/0.49$(\pm0.20)$/0.52$(\pm0.23)$  &0.75 $(\pm0.10)$/0.86$(\pm0.07)$/0.66$(\pm0.12)$ & 0.95/0.95/1\\
			\end{tabular}\\\vspace{1em}}
		\caption{Résultats de MoCaNA face aux finalistes de l'ANAC~2014 (sans~élagage/ élagage~fixe/ élagage~variable)}
		\label{results}
	\end{sidewaystable}
	
	La \Cref{results} montre les résultats de notre agent face aux finalistes de l'ANAC~2014, décrits par \cite{Fukuta2016}, ordonnés selon leur résultats à l'ANAC~2014 dans la catégorie utilité individuelle. Les scores représentent l'utilité moyenne de notre agent et de son adversaire moyenné sur 20 négociations, 10 avec chaque profil. Plus le score est haut, plus l'utilité de l'agent est élevée et mieux l'agent a réussi la négociation. Le nombre suivant le score précédé du signe $\pm$ est l'écart type de la série. L'utilité moyenne est calculée en ne prenant en compte que les négociations qui ont réussi. La dernière colonne représente le taux d'accord atteint par les agents, c'est à dire le nombre de fois où les agents parviennent à un accord divisé par le nombre total de négociations.
	
	Le taux d'accord très faible peut être expliqué par notre stratégie, qui ne s'accorde pas du tout à la distance à la borne, puisqu'elle ne la prend pas en compte. Cette caractéristique l'empêche de ressentir la pression du temps (en anglais \textsl{time pressure}) qui est utilisée par les autres agents pour décider s'ils concèdent beaucoup ou non. Plus de 50\% des négociations se soldent par un échec (contre 45\% pour les autres agents). 
	
	Les deux versions de l'élagage amènent une amélioration significative. Avec l'élagage par une valeur fixe, notre agent ne fait et n'accepte que des propositions pour lesquelles il recevrait une utilité de 0.8 au moins. Cela a un effet important sur le résultat des négociations. Celles qui se terminent permettent à notre d'agent d'obtenir une bonne utilité, mais elles sont moins nombreuses que dans le cas précédent, en particulier en ce qui concerne les agents avec lesquels notre agent avait déjà un taux d'accord faible.
	
	L'autre méthode d'élagage est moins dure pour les adversaires, puisque l'élagage repose sur les propositions qu'ils font tout au long de la négociation. L'utilité de l'agent, en revanche, est bien meilleure que dans le cas sans élagage. Elle l'est moins, en revanche, que dans le cas avec un élagage fixe de $0.8$. Notre agent bat la majorité de ses compétiteurs.
	
	Quelle que soit la technique d'élagage, elle permet aux agents de ne pas explorer les branches inintéressantes, et donc de rechercher les meilleures solutions parmi les solutions acceptables. Elles sont en cela meilleure que la pondération \textsl{a posteriori}.
	
	\subsubsection{Contexte sans borne}
	
	En ce qui concerne la négociation avec le \textsl{Random Walker}, MoCaNA semble beaucoup mieux s'en sortir, même sans élagage. Il obtient un score élevé, 0.703$(\pm0.062)$ contrairement au \textsl{Random Walker}\footnote{Cet agent est décrit dans \cite{Baarslag2013} et fait des propositions aléatoires.}, qui obtient un score moyen de 0.378$(\pm0.077)$.  ce qui confirme que notre agent est gêné par la présence de la borne, et le fait qu'il ne s'appuie pas dessus.
	
	Notre agent est donc un bon négociateur dès lors qu'on ne lui impose pas de borne. La présence de cette dernière coupe la négociation avant qu'il ne puisse arriver à un accord avantageux pour lui. Les négociation réussissant (moins de 50\%), sont donc celles pour lesquelles il a fait le plus de concessions, ce qui a pour effet de lui conférer une utilité moindre que celle de son adversaire.
	\section{Conclusion}
	\label{conclusion}
	Dans cet article, nous avons présenté un agent de négociation automatique capable de négocier dans un contexte où les agents ne se fixent pas de borne, ni temporelle ni en termes de nombre de tours et où le domaine de négociation peut être continu. Après avoir décrit la négociation sous forme de jeu extensif, nous avons décrit une stratégie d'offres s'appuyant sur une version de MCTS et sur deux méthodes de modélisation de l'adversaire. Pour l'une d'entre elles, la régression de processus gaussiens, dont nous avons testé plusieurs \textsl{kernels} et retenu le meilleur. Nous avons proposé deux variantes à notre stratégie d'offre s'appuyant sur des élagages des branches.
	
	Aucun agent n'est conçu pour négocier dans un contexte aussi réaliste que celui pour lequel MoCaNA a été conçu. Nous l'avons donc comparé aux finalistes de l'ANAC 2014, dont le contexte est le plus proche de celui pour lequel notre agent est conçu et à un RandomWalker. Les expérimentations ont montré qu'il est difficile d'obtenir un haut taux d'accord pour un agent n'ayant pas conscience de la \textsl{time pressure}. Dans les cas où un accord est trouvé, l'élagage améliore grandement la performance de MoCaNA et lui permet de battre la majorité de ses adversaires, au prix d'une baisse du taux d'acceptation.
	
	Les perspectives de ces travaux comprennent l'élargissement du spectre de domaines sur lequel MoCaNA est capable de négocier, notamment à tester des domaines de négociations formés d'attributs catégoriels uniquement, continus uniquement, et mixtes. Un autre type de protocoles à tester réside dans les protocoles multilatéraux, notamment des protocoles $1{:}n$ de type enchères ou $n{:}m$ de type \textsl{many-to-many bargaining}. Une autre perspective consisterait à concevoir une version de l'ANAC sans borne. Il serait enfin intéressant de voir dans quelle mesure MoCaNA pourrait intégrer des informations supplémentaires telles que la borne, des connaissances sur la nature des attributs afin d'en tirer parti lorsque cela est possible.
	
	Il est aussi possible d'améliorer MoCaNA en se concentrant sur MCTS. Il serait par exemple intéressant de voir comment un double élagage, par une valeur fixe et selon les propositions de l'adversaire, influerait sur l'utilité et le taux d'accord de MoCaNA. Une autre amélioration, qui permettrait d'augmenter le nombre de simulations de MoCaNA et donc de diminuer le temps de calcul ou d'augmenter la performance de notre agent serait d'implémenter des algorithmes comme \textsl{All Moves As First} ou la \textsl{Rapid Action Value Estimation}.
	
	\bibliographystyle{plain}
	\bibliography{biblio.bib}
\end{document}